\documentclass{styles/svproc}
\usepackage[utf8]{inputenc}
\usepackage{tabularx}
\usepackage{url}
\usepackage{todonotes}
\usepackage{graphicx}
\usepackage{comment}

\begin{document}
\mainmatter              
\title{Blockchain and Biometrics: A First Look into Opportunities and Challenges}
\titlerunning{Blockchain for Biometrics: A First Look}  
%
\author{Oscar Delgado-Mohatar\inst{} \and Julian Fierrez\inst{} \\
    Ruben Tolosana\inst{} \and Ruben Vera-Rodriguez\inst{}}
\authorrunning{Oscar Delgado-Mohatar, Julian Fierrez, et al.} 
%
\tocauthor{Oscar Delgado-Mohatar, Julian Fierrez, Ruben Tolosana and Ruben Vera-Rodriguez}
\institute{Escuela Politecnica Superior,\\ Universidad Autonoma de Madrid, Madrid, Spain\\
\email{\{oscar.delgado, julian.fierrez, ruben.tolosana, ruben.vera\}@uam.es}}

\maketitle              

\begin{abstract}

Blockchain technology has become a thriving topic in the last years, making possible to transform old-fashioned operations to more fast, secured, and cheap approaches. In this study we explore the potential of blockchain for biometrics, analyzing how both technologies can mutually benefit each other. The contribution of this study is twofold: 1) we provide a short overview of both blockchain and biometrics, focusing on the opportunities and challenges that arise when combining them, and 2) we discuss in more detail blockchain for biometric template protection.

\end{abstract}

\section{Introduction}\label{sec:intro}

Among all current disruptive technologies, both blockchain and biometrics have become a focus of attention in recent years. On the one hand, blockchain technology provides an immutable and decentralized data registry, optionally with the capability of executing distributed secure code. Its origins are linked to Bitcoin cryptocurrency, created in 2009, where it is used for solving an old problem opened since the 80's in the cryptographic community: the design of a distributed algorithm of consensus on economic transactions without the participation or existence of a central authority \cite{Nakamoto}. However, nothing prevents any other digital data from being stored instead of economic transactions. This aspect opens the doors to many different potential applications such as smart energy and grids~\cite{Aggarwal2018,Magnani2018}, healthcare~\cite{GORDON2018224}, and smart devices or digital identity schemes~\cite{Stokkink2018DeploymentOA}, among others.

On the other hand, the aim of biometric technology is to authenticate the identity of subjects through the use of physiological (e.g., face, fingerprint) or behavioral (e.g., voice, handwritten signature) traits~\cite{jain_2016}. Its advantages over traditional authentication methods (e.g., no need to carry tokens or remember passwords, they are harder to circumvent, and provide at the same time a stronger link between the subject and the action or event) have allowed a wide deployment of biometric systems, including large-scale national and international initiatives~\cite{SmartBordersEU2013,DaugmanAadhar2014}.



Combining blockchain and biometrics could potentially have many advantages. As a first approximation, the blockchain technology could provide biometric systems with some desirable characteristics such as \textbf{immutability}, \textbf{accountability}, \textbf{availability} or \textbf{universal access}:
\begin{itemize}
    \item By definition, a blockchain guarantees the \textit{immutability} of the registers it stores\footnote{Strictly speaking, a blockchain is not a tamper-proof mechanism but tamper-evident.}, which could be used by a biometric system to build a secure template storage.
    \item Derived from previous property, a blockchain increases the \textit{accountability} and \textit{auditability} of the stored data, which can be very useful to demonstrate to a third party (e.g., a regulator) that the biometric patterns have not been modified.
    \item For last, a (public) blockchain also provides complete \textit{availability} and \textit{universal access} for any user.
\end{itemize}

\begin{table}[t]
\caption{Blockchain / biometrics mutual benefits}
\label{tab:benefits}
\begin{center}
\begin{tabularx}{\textwidth}{m{5cm} X}
\hline
Blockchain to biometrics &  Immutability \\
                         & Accountability \\
                         & Availability \\
                         & Universal access \\
\hline
Biometrics to blockchain  & More secure digital identity models \\
                          & New use cases (e.g., \textit{smart devices}) \\
                          & Biometric wallets\\[2pt]
\hline
\end{tabularx}
\end{center}
\end{table}

Additionally, the integration of biometric technology would be very beneficial for blockchains too. Among many other new use cases, biometrics could greatly improve the current distributed digital identity schemes based on blockchain. Another interesting application of biometrics to blockchain is related to \textit{smart devices}. A smart device is any digital or physical asset with access to a blockchain that can perform actions and make decisions based on the information stored there. For example, a car could be fully managed (rented or bought) through a smart contract. However, an adequate identification of the user is not fully solved yet. In this case, an authentication protocol based on biometrics could significantly raise the current security level. Table \ref{tab:benefits} overviews the mutual benefits of blockchain and biometrics.

The main contributions of this paper can be summarized as follows:
\begin{itemize}
    \item We provide a short overview of both blockchain and biometrics, focusing on the opportunities and challenges that arise when combining them.
    \item We discuss in more detail an architecture for biometric template protection based on blockchain.
\end{itemize}

The remainder of the paper is organized as follows. In Section \ref{sec:blockchain} a description of the most relevant features of blockchain for biometric technologies is provided. In Section \ref{sec:blockchain_for_biometrics}, we first analyze the challenges and limitations of the technologies to finally discuss blockchain for biometric template protection. Finally, Section \ref{sec:conclusions} draws the final conclusions and points out some lines for future work.

\section{Blockchain Basics}\label{sec:blockchain}


\begin{table}[t]
\caption{Characterization of main blockchain platforms}
\label{tab:main_blockchains}
\begin{center}
\begin{tabularx}{\textwidth}{m{3cm} X X X}
\hline
\rule{0pt}{12pt} \textbf{Blockchain type} &  \textbf{Public} & \textbf{Consortium} & \textbf{Private} \\[2pt]
\hline \hline
Governance  &  No centralized \newline management & Multiple \newline organizations & Tipically, single \newline organization \\
Access control  &  Permissionless & \multicolumn{2}{c}{Permissioned} \\
Participants  &  Anonymous & \multicolumn{2}{c}{Identified, trusted} \\
Main platform  &  Bitcoin, Ethereum & Quorum, Parity & Hyperledger \\
Consensus algorithm & PoW / PoS & \multicolumn{2}{c}{Voting or multi-party} \\
 & & \multicolumn{2}{c}{consensus algorithms (PoS /PoA)} \\
Transaction \newline confirmation time & Long (minutes) & Short & Short \\
Data privacy & No & Optional & Yes \\
Smart contracts \newline support & Very limited & Yes & Yes \\
Cryptocurrency & BTC & ETH & - \\
\hline
\end{tabularx}
\end{center}
\end{table}

\subsection{Overview}\label{subsec:overview}
Essentially, a blockchain is a decentralized public ledger of all data and transactions that have ever been executed in the system \cite{swan2015blockchain}. These transactions are recorded in blocks that are created and added to the blockchain in a linear, chronological order (immutable). Each participating node in the network has the task of validating and relaying transactions, and has a copy of the blockchain.

However, since its initial application to Bitcoin cryptocurrency, the original idea of a universal and public blockchain has greatly evolved into new architectures, based on different access control schemes or consensus algorithms.

According to the first criteria, blockchains can be categorized as: 1) public, 2) consortium, and 3) private blockchains (see Table \ref{tab:main_blockchains}). Essentially, public blockchains are permissionless schemes, designed with a built-in economic incentive for allowing anonymous and universal access. Consortium blockchains, on the other hand, are permissioned, partly private and semi-decentralized architectures, specially targeted for scenarios with a small number of participants. Last, private blockchains are specially indicated in applications where users must be fully identified and trusted. This application environment for private blockchain makes more straightforward the incorporation of biometrics compared to public and consortium blockchains. Anyway, in all three types of blockchains (public, consortium, and private) further research and new security architectures are needed to deliver the full potential of the excellent synergies between blockchain and biometrics.


Blockchains can also use different consensus algorithms, some of which allow greater efficiency and faster transactions completion time. Therefore, the most appropriate type of blockchain depends on the specific use case.

\subsection{Smart Contracts}\label{subsec:applications}

The term \textit{smart contract} dates back to 1996, long before the creation of Bitcoin and blockchain, and was first introduced by Nick Szabo \cite{Szabo1996}. A smart contract is, essentially, a piece of code executed in a secure environment that controls digital assets. Examples of these secure environments include regular servers controlled by ``trusted parties'', decentralized networks (blockchains), or servers with secure hardware (SGX) \cite{Karande2017,Kucuk2016}.

Many public blockchains support the execution of smart contracts, but the most reliable, secure, and used is, without doubt, Ethereum \cite{Dannen2017}. Ethereum could be considered as a distributed computer, with capability to execute programs written in Turing-complete, high-level programming languages. These programs are no more than a collection of pre-defined instructions and data that has been recorded at a specific address of a blockchain.

For biometric purposes, a smart contract running in a blockchain can assure a semantically correct execution. However, the consensus algorithms necessary to provide this security in public blockchains have an associated economic cost, which will be analyzed in next section.


\section{Blockchain for Biometrics}\label{sec:blockchain_for_biometrics}

\subsection{Challenges and Limitations}\label{subsec:challenges}
Despite the new opportunities already described in previous sections, the combination of both blockchain and biometric technologies is not straightforward due to the limitations of the current blockchain technology. Among them, it is important to remark: 1) its transaction processing capacity is currently very low (around tens of transactions per second), 2) its actual design implies that all system transactions must be stored, which makes the storage space necessary for its management to grow very quickly, and 3) its robustness against different types of attacks has not been sufficiently studied yet.

We now detail the limitations of blockchain public networks for the deployment and operation of biometric systems.

\begin{itemize}
\label{sec:coste}
\item \textbf{Economic cost of executing smart contracts:} In order to support smart contracts in blockchains (like Ethereum), and to reward the nodes that use their computing capacity to maintain the system, each instruction executed requires the payment of a fee in a cryptocurrency (called gas). Simple instructions (such as a sum) cost 1 gas, while others can cost significantly more (e.g., the calculation of a SHA3 hash costs 20 gas). On the other hand, the storage space is especially expensive (around 100 gas for every 256 bits). Therefore, one of the first research problems would be minimizing the cost of running a biometric system (totally or partially) in a blockchain, and how efficiently smart contracts involving biometrics could be coded.

\item \textbf{Privacy:} By design, all operations carried out in a public blockchain are known by all the participating nodes. Thus, it is not possible to directly use secret cryptographic keys, as this would reduce the number of potential applications.
Regarding privacy in public blockchains, three main layers are considered in general: 1) participants, 2) terms, and 3) data. The first one ensures participants to remain anonymous both inside and outside of the blockchain. This is achieved with cryptographic mechanisms like ring signatures, stealth addresses, mixing, or storage of private data off-chain. Second, privacy of terms keeps the logic of the smart contracts secret, by using range proofs or Pedersen commitments. Last, and the most important for biometrics, the data privacy layer goal is to keep transactions, smart contracts, and other data such as biometric templates, encrypted at all times, both on-chain and off-chain. The cryptographic tools used include zero-knowledge proofs (ZKP) and zk-SNARKS, Pedersen commitments, or off-chain privacy layers like hardware-based trusted execution environments (TEEs). However, the application of these cryptographic tools are still very limited for blockchains. For example, Ethereum just included at the end of 2017 basic verification capabilities for ZKPs. More advanced cryptographic tools have been only developed to target special cases like Aztec~\cite{Williamson2018} or ZK range proofs~\cite{Koens2018}. In addition, it should be noted that ZKP transactions would be still expensive and computationally intensive ($\sim$~1,5M gas/verification).

\item \textbf{Processing capability:} Another important limitation is related to its processing capability. Ethereum, for example, is able to run just around a dozen transactions per second, what it could be not enough for some scenarios. Additionally, there is a minimum confirmation time before considering that the transaction has been properly added to the blockchain. This time can oscillate among different blockchains, from tens of seconds to minutes, reducing its usability for biometric systems.

\item \textbf{Scalability:} This is one of the main handicaps of the technology from its origins as, theoretically, all nodes of the blockchain network must store all blocks of the blockchain network. Currently, the size of the public blockchains (Bitcoin and Ethereum) is around 200GB, and it is growing very fast. This can be a problem for some application scenarios such as the Internet of Things (IoT).

\item \textbf{Security:} As novel technology, blockchain security characterization is still a work in progress. Among all possible attacks, it is worth mentioning the attack known as \textit{51\% attack}~\cite{Eyal2014b}. If an attacker gains more than 50\% of the computational capacity of any public or private blockchain, he could reverse or falsify transactions. This attack applies even to blockchain with consensus algorithms not based in proof-of-works schemes, like PoS or PoA, typically used in private or consortium topologies. However, the main security problems suffered to date by blockchains are mainly related to programming errors, e.g., the DAO attack happened in 2016, which put at risk the whole Ethereum ecosystem~\cite{Atzei2017}.

\end{itemize}

\subsection{Blockchain for Biometric Template Protection}\label{subsec:proposed_approach}

Biometric systems have for long been known to be vulnerable to certain physical \cite{hadid15SPMspoofing} and software attacks \cite{barrero13PRLmultimodalAttack}. Physical attacks to the biometric sensor can be overcome to some extent with presentation attack detection techniques \cite{2019_Hand_Fierrez}. On the other hand, an important group of software attacks can be prevented using biometric template protection techniques, but the state-of-the-art there \cite{2017_Access_HEmultiDTW_Marta,2017_PR_multiBtpHE_marta} is still improvable in many ways \cite{2018_INFFUS_MCSreview2_Fierrez}.

\begin{figure}[tb]
\caption{Main security vulnerabilities of biometric systems and biometric template protection based on blockchain.}
\centering
\vspace{0.75cm}
\label{fig:proposal}
\includegraphics[width=\textwidth]{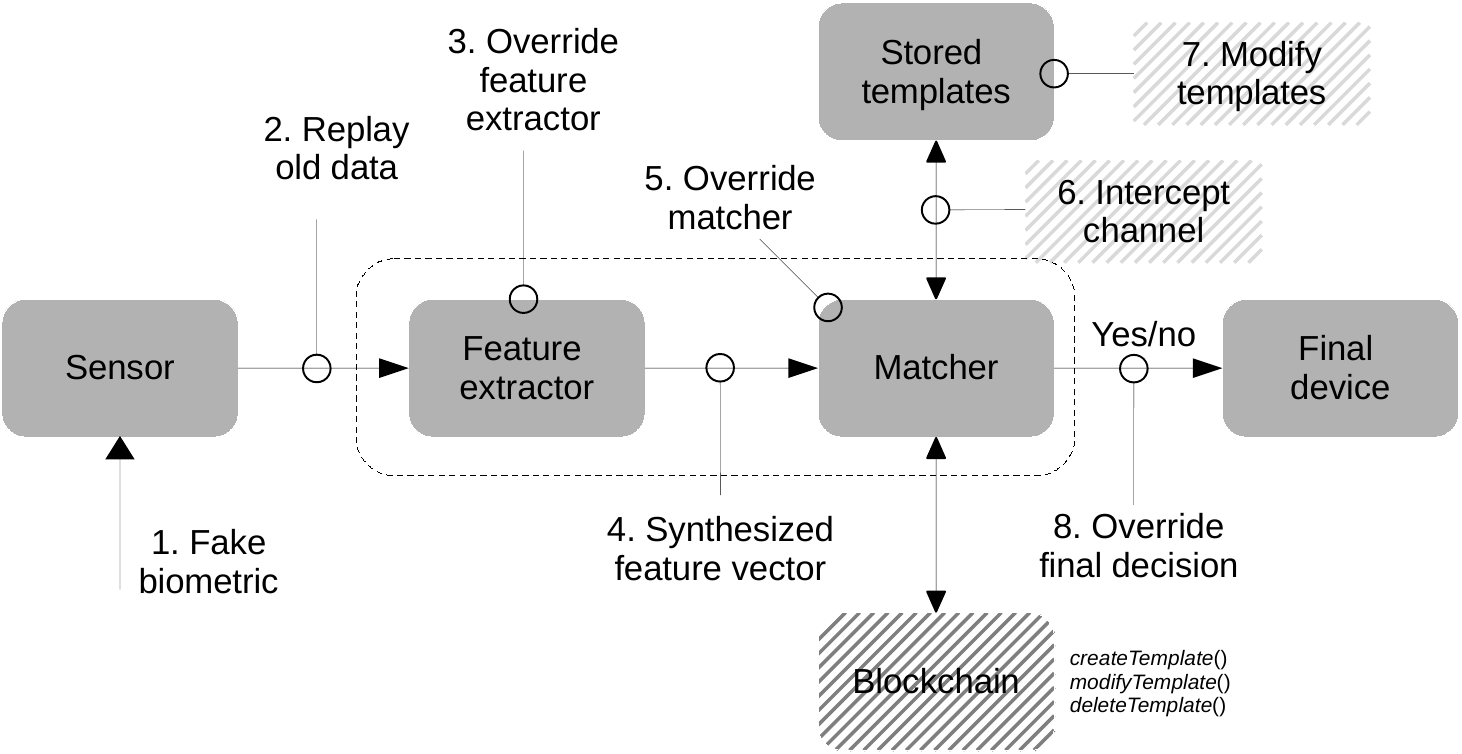}
\end{figure}

Figure~\ref{fig:proposal} depicts the typical stages of a biometric system (in solid gray), all possible points where a biometric system can be attacked, and a representation of biometric template protection based on blockchain (stripped block). By substituting the traditional template store by a blockchain, the security level of the resulting biometric system is significantly increased. If correctly implemented, attacks number 6 (channel interception) and 7 (templates modification) and are no longer possible.



This design provides some advantages:
\begin{itemize}
    \item The modifications to the existing biometric systems are kept to a minimum, so the usual biometric techniques and algorithms (e.g., feature extraction and matching) can be used normally.
    \item Since the biometric process is performed off-chain, this architecture avoids the scalability problems of public blockchains (except in a massive batch of user registration).
    \item No need to use complex smart contracts, which facilitates development and reduces execution costs. Smart contracts do not implement biometric ``logic'', but only the minimum necessary functions to manage the storage of templates (creation, modification, etc.)
\end{itemize}

However, as stated before, storage space in blockchains is specially expensive compared to computation, in order to discourage its abusive use. As an example, for current Ether price (around 140\$ at time of writing, February 2019), a 1KB fingerprint template would cost around 0.00093\$ to be stored in Etherum. In any case, blockchains do not usually store data directly, but use distributed storage platforms like IPFS \cite{Benet2014}.




\section{Conclusions and Future Work}\label{sec:conclusions}%

Although research on the integration of biometric systems and blockchains is incipient and is taking its first steps, it is undeniable that both technologies have a potential for collaboration and enormous mutual growth.

In this paper we have discussed the main characteristics and limitations of blockchains, especially those that could directly affect the implementation of biometric systems. We have also explored the potential mutual benefits for both technologies, and discussed a first approximation to a combined architecture using blockchain for biometric template protection.

With a view in the future, a key question arises: how many of the biometric processes can be integrated or ported into a blockchain, i.e., done on-chain?. For example, would it be possible to implement a biometric matcher using a smart contract? how? which challenges should be solved to do so?

Due to the current limitations and characteristics of the blockchain technology, a full integration with biometric processes seems very challenging in the short term. However, there are some promising research areas, e.g.,  the use of \textbf{state channels} \cite{Dziembowski2018}, which could drastically reduce costs and improve bandwidth, or the development of new zero-knowledge proofs that would allow a user to be authenticated through biometrics without any of the parties having knowledge of the user's identity.

\section*{Acknowledgements}

Research supported by project TEC2015-70627-R (MINECO/FEDER), UAM-CecaBank chair on Biometrics, and UAM-GrantThornton chair on Blockchain. Ruben Tolosana is supported by a FPU Fellowship from Spanish MECD.

%
%
%
\bibliographystyle{plain}

\begin{thebibliography}{10}

\bibitem{Aggarwal2018}
Shubhani Aggarwal et~al.
\newblock Energychain: Enabling energy trading for smart homes using
  blockchains in smart grid ecosystem.
\newblock In {\em Proc. of SmartCitiesSecurity}, 2018.

\bibitem{Atzei2017}
Nicola Atzei et~al.
\newblock A survey of attacks on ethereum smart contracts {SoK}.
\newblock In {\em Proc. Intl. Conf. on Principles of Security and Trust}.
  Springer, 2017.

\bibitem{Benet2014}
Juan Benet.
\newblock {IPFS - Content Addressed, Versioned, P2P File System}.
\newblock jul 2014.

\bibitem{SmartBordersEU2013}
European Commission.
\newblock {'Smart Borders': for an open and secure EU}, 2013.

\bibitem{Dannen2017}
Chris Dannen.
\newblock {\em Introducing Ethereum and Solidity: Foundations of Cryptocurrency
  and Blockchain Programming for Beginners}.
\newblock Apress, Berkeley, CA, USA, 2017.

\bibitem{DaugmanAadhar2014}
John Daugman.
\newblock 600 million citizens of india are now enrolled with biometric {ID}.
\newblock {\em SPIE Newsroom}, 2014.

\bibitem{Dziembowski2018}
Stefan Dziembowski et~al.
\newblock General state channel networks.
\newblock In {\em Proc. of ACM SIGSAC Conf. on Computer and Communications
  Security}, CCS '18, 2018.

\bibitem{Eyal2014b}
Ittay Eyal and Emin~G\"{u}n Sirer.
\newblock Majority is not enough: Bitcoin mining is vulnerable.
\newblock {\em Commun. ACM}, 61(7):95--102, June 2018.

\bibitem{2018_INFFUS_MCSreview2_Fierrez}
Julian Fierrez et~al.
\newblock Multiple classifiers in biometrics. part 2: Trends and challenges.
\newblock {\em Information Fusion}, 44:103--112, November 2018.

\bibitem{2017_PR_multiBtpHE_marta}
Marta Gomez-Barrero et~al.
\newblock Multi-biometric template protection based on homomorphic encryption.
\newblock {\em Pattern Recognition}, 67:149--163, July 2017.

\bibitem{barrero13PRLmultimodalAttack}
Marta Gomez-Barrero, Javier Galbally, and Julian Fierrez.
\newblock Efficient software attack to multimodal biometric systems and its
  application to face and iris fusion.
\newblock {\em Pattern Recognition Letters}, 36:243--253, January 2014.

\bibitem{2017_Access_HEmultiDTW_Marta}
Marta Gomez-Barrero, Javier Galbally, Aythami Morales, and Julian Fierrez.
\newblock Privacy-preserving comparison of variable-length data with
  application to biometric template protection.
\newblock {\em IEEE Access}, 5:8606--8619, June 2017.

\bibitem{GORDON2018224}
W.J. Gordon and C.~Catalini.
\newblock Blockchain technology for healthcare: Facilitating the transition to
  patient-driven interoperability.
\newblock {\em Computational and Structural Biotechnology Journal},
  16:224--230, 2018.

\bibitem{hadid15SPMspoofing}
A.~Hadid, N.~Evans, S.~Marcel, and J.~Fierrez.
\newblock Biometrics systems under spoofing attack: an evaluation methodology
  and lessons learned.
\newblock {\em IEEE Signal Processing Magazine}, 32(5):20--30, September 2015.

\bibitem{jain_2016}
Anil~K. Jain et~al.
\newblock {50 years of biometric research: Accomplishments, challenges, and
  opportunities}.
\newblock {\em Pattern Recognition Letters}, 79:80--105, 2016.

\bibitem{Karande2017}
Vishal Karande et~al.
\newblock {SGX-Log}: Securing system logs with {SGX}.
\newblock In {\em Proc. of ACM Asian Conf. on Computer and Communications
  Security}, 2017.

\bibitem{Koens2018}
Tommy Koens, Coen Ramaekers, and Cees {Van Wijk}.
\newblock {Efficient Zero-Knowledge Range Proofs in Ethereum}.
\newblock Tech. Report, 2018.

\bibitem{Kucuk2016}
Kubilay~A. K\"{u}\c{c}\"{u}k et~al.
\newblock Exploring the use of {Intel SGX} for secure many-party applications.
\newblock In {\em Workshop on System Software for Trusted Execution}, 2016.

\bibitem{Magnani2018}
Antonio Magnani et~al.
\newblock Feather forking as a positive force: Incentivising green energy
  production in a blockchain-based smart grid.
\newblock In {\em ACM Workshop on Cryptocurrencies and Blockchains for
  Distributed Systems}, 2018.

\bibitem{2019_Hand_Fierrez}
S.~Marcel, M.~Nixon, J.~Fierrez, and N.~Evans, editors.
\newblock {\em Handbook of Biometric Anti-Spoofing - Presentation Attack
  Detection, Second Edition}.
\newblock Springer, 2019.

\bibitem{Nakamoto}
Satoshi Nakamoto.
\newblock {Bitcoin: A Peer-to-Peer Electronic Cash System}, 2008.

\bibitem{Stokkink2018DeploymentOA}
Quinten Stokkink and Johan~A. Pouwelse.
\newblock Deployment of a blockchain-based self-sovereign identity.
\newblock {\em CoRR}, abs/1806.01926, 2018.

\bibitem{swan2015blockchain}
Melanie Swan.
\newblock {\em Blockchain: Blueprint for a New Economy}.
\newblock O'Reilly, 2015.

\bibitem{Szabo1996}
Nick Szabo.
\newblock {Smart Contracts: Building Blocks for Digital Markets}, 1996.

\bibitem{Williamson2018}
Zachary~J. Williamson.
\newblock {The AZTEC Protocol}.
\newblock Tech. Report, 2018.

\end{thebibliography}

\end{document}